\documentclass[sensors,article,accept,pdftex,moreauthors]{Definitions/mdpi} 
\usepackage[normalem]{ulem}
\newif\ifrev
\revtrue 
\ifrev
\newcommand{\revminor}[1]{\textcolor{black}{#1}}
\else
\newcommand{\rev}[1]{#1}
\newcommand{\revtech}[1]{#1}
\newcommand{\rmtexto}[1]{}
\fi
\newcommand{\rev}[1]{#1}
\newcommand{\revtech}[1]{#1}
\usepackage[none]{hyphenat}
\usepackage{makecell}

\firstpage{1} 
\pubvolume{1}
\issuenum{1}
\articlenumber{0}
\pubyear{2022}
\copyrightyear{2022}
\externaleditor{Academic Editors: Adam M. Kawalec, Marta Walenczykowska and Ksawery Krenc} 
\datereceived{5 July 2022} 
\dateaccepted{7 September 2022} 
\datepublished{} 
\hreflink{https://doi.org/} 

\Title{Non-Ionizing Radiation Measurements for Trajectography~Radars}

\TitleCitation{Non-Ionizing Radiation Measurements for Trajectography Radars}

\Author{J. 
 Marcos Leal {Barbosa} 
 Filho $^{*}$\orcidA{}, Millena M. de M. Campos \orcidB{}, Daniel L. Flor
\orcidC{}, William S. Alves
\orcidD{}, \mbox{Adaildo G. D'Assunção \orcidE{},} Marcio E. C. Rodrigues \orcidG{} and Vicente A. de Sousa Jr. \orcidH{}}

\AuthorNames{J. Marcos Leal B. Filho, Millena M. de M. Campos, Daniel L. Flor, William S. Alves, Adaildo G. D'Assunção, Marcio Eduardo C. Rodrigues, Vicente A. de Sousa Jr.}

\AuthorCitation{Barbosa Filho, J.M.L.; 
 Campos, M.M.d.M.; Flor, D.L.; Alves, W.S.; D'Assunção, A.G.; Rodrigues, M.E.C.; de Sousa, 
 V.A.,  Jr.}

\address[1]{%
{Department} 
 of Communications Engineering, Federal University of Rio Grande do Norte, \mbox{Natal 59078-970, Brazil;}  millenacmps@gmail.com (M.M.d.M.C.);  daniel.flor.087@ufrn.edu.br (D.L.F.); wsalves0409@gmail.com (W.S.A.); adaildo@ct.ufrn.br (A.G.D.); marcio.rodrigues@ufrn.br (M.E.C.R.); vicente.sousa@ufrn.br (V.A.d.S.J.)}

\corres{\hangafter=1 \hangindent=1.05em \hspace{-0.82em} Correspondence: marcoslealufrn@gmail.com}

\abstract{This work presents a Non-Ionizing Radiation (NIR) measurement campaign and proposes a specific measurement method for trajectography radars. This kind of radar has a high gain narrow beam antenna and emits a high power signal. Power density measurements from a C-band trajectography radar are carried out using bench equipment and a directional receiving antenna, instead of the commonly used isotropic probe. The measured power density levels are assessed for compliance test via comparison with the occupational and general public exposure limit levels of both the International Commission on Non-Ionizing Radiation Protection (ICNIRP) and the Brazilian National Telecommunication Agency (Anatel). The limit for the occupational public is respected everywhere, evidencing the safe operation of the studied radar. However, the limit for the general public is exceeded at a point next to the radar’s antenna, showing that preventive measures are needed.}


\begin{document}


\section{Introduction}

In these last years, the~number of satellites, spacecrafts, astronaut crews, and~other technological artifacts launched into space has grown significantly. In~launch campaigns, for~the tracking of rockets and their payloads, there is an essential element: the trajectography radar. It performs a Single-Target-Tracking (STT)~\cite{lac2001} by automatically tracking a specific target, recording its trajectory accurately and in real time. Among~its main applications stand out the tracking of rockets, missiles, satellites and space~debris.

Two trajectography radars are used at a Brazilian Air Force launch base named Centro de Lançamento da Barreira do Inferno (CLBI), located in Parnamirim city, state of Rio Grande do Norte. This base performs and provides support for launches, tracking activities and research. One of these activities is the tracking of rockets launched by the Guiana Space Centre at Kourou, in~French Guiana, in~cooperation with the European Space Agency (ESA). For~the tracking activities, CLBI has the Adour and Béarn trajectography radars and a telemetry~station. 

The radars are French-made, C-band operated, and~designed to track targets over long distances. For~that, the~Béarn radar has a high gain (about 44 dBi) narrow beam (about 0.92° half power beam width) antenna, a~high-sensitivity receiver and a transmitter capable of delivering high-power pulsed electromagnetic signals, reaching levels around 1 kW of average power and 1 MW of peak power~\cite{THOMSON-CSF}.


Trajectography radars emit high power of non-ionizing radiation (NIR), concentrated in a narrow beam. This set of specifications constitutes a quite unusual radiofrequency emission scenario when compared to the general telecommunication broadcasts, such as TV signal transmission, personal mobile service coverage and satellite communication. Radars’ high-power emissions cause concerns about the possible danger to human health, especially in public areas. For~health safety reasons, there is interest in measuring the NIR levels that people who work daily with the radar and those who visit it are exposed to. From~the measured values, it is possible to make a comparative analysis with the human exposure limits to NIR specified by the International Commission for the Protection of Non-Ionizing Radiation (ICNIRP) \cite{ICNIRP-2020} and also by the Brazilian National Telecommunication Agency (Anatel) \cite{ANATEL_ATO_458}. Then, measures can be taken to prevent and control this~exposure.

In this work, field measurements are performed to obtain the power density values of Béarn’s emissions, which operate with a higher power than Adour radar and are located only 690 m from a public space museum. This is one of the chosen measurement points, as~it receives daily visits from children and teenagers from local schools and tourists from various parts of Brazil and the world. The~objective is to measure only radar emissions, avoiding capturing signals from other sources. For~that, a~specific setup for trajectography radars was proposed. Up~to now, there are no records of RNI levels in CLBI’s~area.


\subsection{Related~Works}


A survey was carried out on the Scopus database, using the following keywords related to the matter: ``radar'' and ``non-ionizing'' and (``radiation'' or ``NIR'' or ``electromagnetic field'' or ``EMF'' or ``measurement''), looking in the title, abstract and keywords fields. Considering the last 50 years, 93 publications were found. Among~them, 46 publications have no direct relationship with radars; 35 are related to handheld radars mostly used for cancer detection and other health analyses, but~they are not related to NIR \mbox{studies~\cite{bha2022,men2020,owd2020};} 9 are radar and NIR related, but~measurements were not performed~\cite{pel2018, yak2011, gol1995}; and only 3 contributions carry out studies on NIR radar measurements (air traffic control and ship-board navigation radars) \cite{jos2012,mar2013,hal2015}. One of the reasons for the small number of publications in this area is the restricted access to this type of radar. There were no studies found about measuring non-ionizing radiation due to trajectography radars, evidencing that probably this is the first work to do~so.


References~\cite{jos2012,hal2015} are focused on radars and clearly describe the setup, so they were chosen to be summarized next. In~\cite{jos2012} a measurement campaign is carried out involving 14 types of air traffic control systems, including a Weather Radar (WR) and three types of aircraft tracking radars. Two of them are the Primary Surveillance Radar (PSR) and the Secondary Surveillance Radar (SSR) used together for long-range scans. The~third type is the Surface Radar (SR) which is a short-range radar, often placed in a control tower, used to scan the lower layers of the air and the movement of planes on the ground. The~PSRs are L-band and S-band; SSRs are L-band; the SRs are X-band and Ku-band; and the WR is C-band. These types of radars continuously rotate their antennas horizontally, so the NIR levels are not constant. For~this measurement, a~Rohde and Schwarz HF907OM omnidirectional antenna for large frequency bands (800 MHz--26.5 GHz) was used in combination with a Rohde and Schwarz spectrum analyzer of the FSEM type. The~results show that no limits established by ICNIRP were exceeded, both for the general and occupational~public.  

In~\cite{hal2015} NIR measurements are performed aboard a marine vessel with seven different transmitters. Two of them are X-band and S-band radars, used for navigation purposes. For~that, a~NARDA NBM 520 with an EF 1891 probe (frequency range 3 MHz--18 GHz) are used. The~probe is an isotropic one, providing non-directional measurements. A~method to collect the measured data is proposed, considering the different sources transmitting at different times. Considering ICNIRP specifications, both general and occupational public limits are respected for all measured locations in the ship when the transmitters are operating separately. However, when both radars, satellite communication, automatic identification system navigation and VHF radio are operating at the same time, the~limits for the general population are exceeded on the bridge~roof.

\revtech{Radar NIR measurements are generally carried out with isotropic probes, as~performed in~\cite{jos2012,hal2015}. This is because most radars have antennas that work continuously in rotation or scanning mode, unlike trajectography radar which points the antenna in the chosen direction and at any time. With~this type of probe, it is impracticable to measure the power density of signals exclusively from the radar, unless~it is in an isolated area and it's the only transmitting source, which is generally not the case for ground radars. CLBI's radars, for~example, receive interfering third harmonic signals from mobile communication base stations located in different positions around the radars. To~deal with this situation, considering the high directivity of trajectography radar antennas for accurate tracking of unique targets, a~specific measurement method for this type of radar is proposed here.}

This work is organized as follows. Section~\ref{sec_regula} presents the Anatel Act No. 458 and the national and international established limits regarding human exposure to NIR. Next, Section~\ref{sec_mat_met} introduces the materials and methods, presenting \rev{the proposed method, the~measurement scenario and setup}. Then, Section~\ref{sec_results} presents and discusses the results and comparisons of the measurements to the ICNIRP and Anatel limit values. Finally, Section~\ref{sec_conclu} presents~conclusions.

\section{Regulation of Human Exposure Limits to Non-Ioninzing~Radiation}
\label{sec_regula}

Non-ionizing radiation is a kind of electromagnetic radiation that, as~opposed to ionizing radiation, does not cause an electron withdrawal from molecular structures. Nevertheless, the~incidence of NIR on a material produces thermal and non-thermal effects. The~thermal effects are a consequence of the increased vibration in the material’s molecules caused by the NIR. Concerning the non-thermal effects, there is no evidence that they pose a danger to human health, particularly with regard to long-term exposure~\cite{wol2003}. 

The Electric and Magnetic Fields (EMF) project of the World Health Organization (WHO) has been developing a program that, among~other objectives, evaluates the scientific literature and reports on health effects to facilitate the development of internationally acceptable standards for exposure to EMF. In~addition, this program intends to advise national authorities, other institutions, the~general public and workers about any risks arising from exposure to EMF and any necessary mitigation actions~\cite{WHO_EMF}.

In Brazil, law No. 11934, of~5 May  2009, 
 establishes that the definition of exposure limits to non-ionizing radiation must follow WHO recommendations, which adopted ICNIRP numbers~\cite{ICNIRP-1998}. Currently, these limits are defined by Anatel Act No. 458, of~24~January 2019~\cite{ANATEL_ATO_458}. Compliance verification is based on field measurements so that preventive procedures can be adopted to avoid the impact on human health~\cite{jmoe-2021,Marcio_2015,Marcio_2013}. Anatel Act No. 458 regulates the limits of human exposure to electric, magnetic and electromagnetic fields in the radiofrequency range between 8.3 kHz and 300 GHz, generated by radio-communication stations and user terminals. The~regulation also defines evaluation methods for exposure to non-ionizing radiation and procedures to be followed for the licensing of~stations. 

Different values of exposure limit are defined to the occupational and to the general population. Occupational exposure refers to that in which people are exposed as a result of their professional activity, as~long as they are aware of the potential exposure, and~they can exercise control over their permanence in the place or adopt preventive~measures.

Limit values are the same for the entire range from 2 GHz to 300 GHz, where the Béarn radar operates. For~the general population, the~equivalent plane wave density should be less than 10 W/m² and, for~the occupational public, 50 W/m². Other limits are shown in Table~\ref{tab_pop_exp}~\cite{ANATEL_ATO_458}. 

\begin{table}[H]
    \caption{Population 
 exposure limits to Radio Frequency---Electromagnetic Fields (RF-EMF) for the 2~GHz to 300 GHz~range.}
    \label{tab_pop_exp}
\setlength{\cellWidtha}{\textwidth/4-2\tabcolsep-0in}
\setlength{\cellWidthb}{\textwidth/4-2\tabcolsep-0in}
\setlength{\cellWidthc}{\textwidth/4-2\tabcolsep-0in}
\setlength{\cellWidthd}{\textwidth/4-2\tabcolsep-0in}
		\begin{tabularx}{\textwidth}{>{\centering\arraybackslash}m{\cellWidtha}>{\centering\arraybackslash}m{\cellWidthb}>{\centering\arraybackslash}m{\cellWidthc}>{\centering\arraybackslash}m{\cellWidthd}}
			\toprule
	  \multicolumn{1}{l}{\footnotesize \textbf{Population}} &
	  \multicolumn{1}{l}{\footnotesize \textbf{\makecell{Electric Field \\Intensity (V/m)}}} &
	  \multicolumn{1}{l}{\footnotesize \textbf{\makecell{Magnetic Field \\ Intensity (A/m)}}} &
	  \multicolumn{1}{l}{\footnotesize \textbf{\makecell{Equivalent Plane \\Wave Density (W/m$^2$)}}} \\ \hline
	    \footnotesize General            & \footnotesize 61   & \footnotesize 0.16       &  \footnotesize~10 \\
	    \footnotesize Occupational            & \footnotesize 137   & \footnotesize 0.36       & \footnotesize~50 \\
		\bottomrule
		\end{tabularx}
\end{table}

The power density limits are related to the average power from the emissions. However, the~peak power level from the radar emission (1 MW) is much higher than its average power. Although~there is not much information about the relationship between biological effects and peak values of pulsed fields, it is suggested that, for~frequencies exceeding 10 MHz, the~equivalent plane wave density, averaged over the pulse width, should not exceed 1000 times the reference levels~\cite{ICNIRP-2020}. Since the peak power of the radar signal is exactly 1000 times the average power value, it is within the suggested limit. Therefore, our measurements are focused on the average power~density.

\section{Materials and~Methods}
\label{sec_mat_met}

\revminor{The unique characteristics of trajectography radars require an appropriate method and setup for NIR measurements, which are presented below, as~well as the measurement scenario.}

\subsection{Proposed~Method}

\begin{figure}[H]
     \includegraphics[width=1\linewidth]{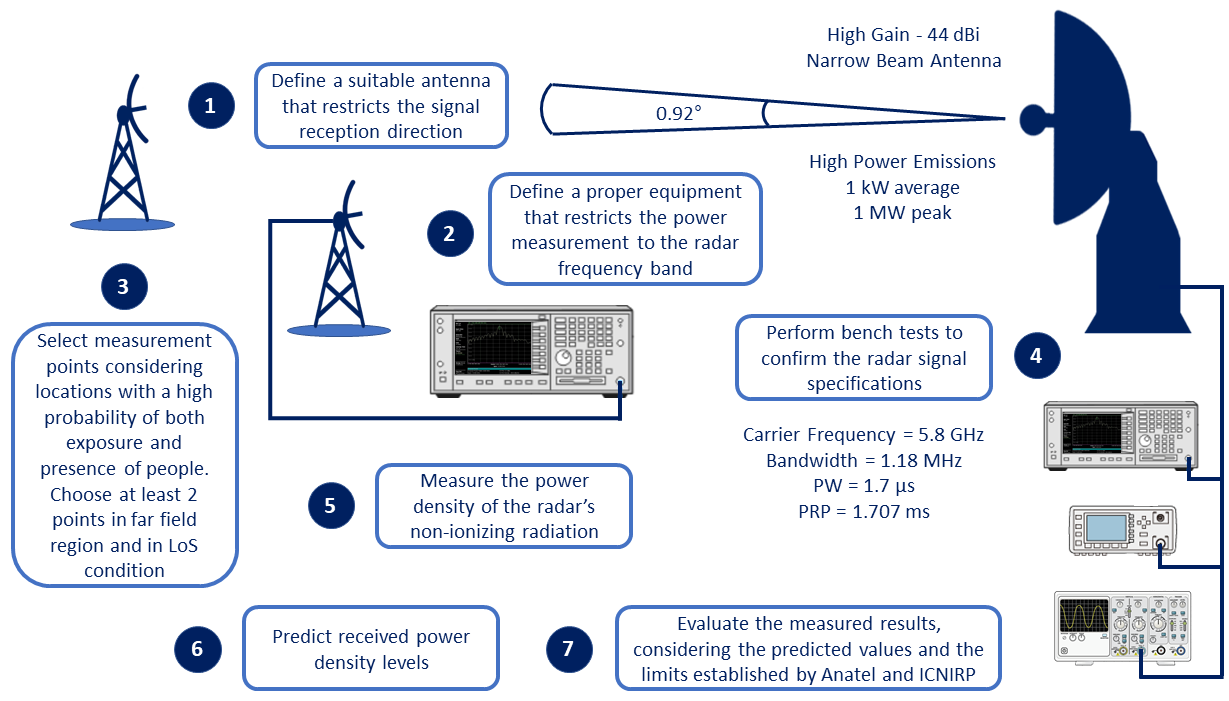}
      \caption{Flowchart of non-ionizing radiation measurement methodology for trajectography~radars.}
     \label{flowchart}
\end{figure}

\revtech{Since the goal of this paper is to measure NIR levels exclusively from the trajectography radar, specific actions and instruments are defined, considering its narrow band (1.18 MHz) signal spectrum and its high gain (about 44 dBi) narrow beam (about 0.92° half power beam width) antenna. A~summary of the proposed steps is presented in Figure~\ref{flowchart} and detailed~below.}

\begin{enumerate}
\item \revtech{Define a suitable receiving antenna for the setup, allowing the restriction of the reception direction, considering only the direction of the radar antenna;}

\item \revtech{Define proper equipment for measuring received power in the frequency domain to restrict the reception of other signals outside the radar frequency band;}

\item \revtech{Select measurement points considering two main criteria: locations with a high probability of both exposure and presence of people. Among~them, choose at least 2 measurement points with Line-of-Sight (LoS) and in the far-field region, in~order to evaluate and compare the measurement results with more precise predictions, using free space propagation equations;}

\item \revtech{Perform bench tests to confirm the radar signal specifications and the adequate performance of the measurement equipment;}

\item \revtech{Prepare the setup and perform the average power density measurements, during~6 min, according to Anatel and ICNIRP instructions~\cite{ANATEL_ATO_458, ICNIRP-2020};}

\item \revtech{Predict received power density levels;}

\item \revtech{Evaluate the measured results, considering the predicted values and the limits established by Anatel and ICNIRP~\cite{ANATEL_ATO_458, ICNIRP-2020}.}

\end{enumerate}

\subsection{Measurement~Scenario}

\begin{figure}[H]
     \includegraphics[width=0.75\linewidth]{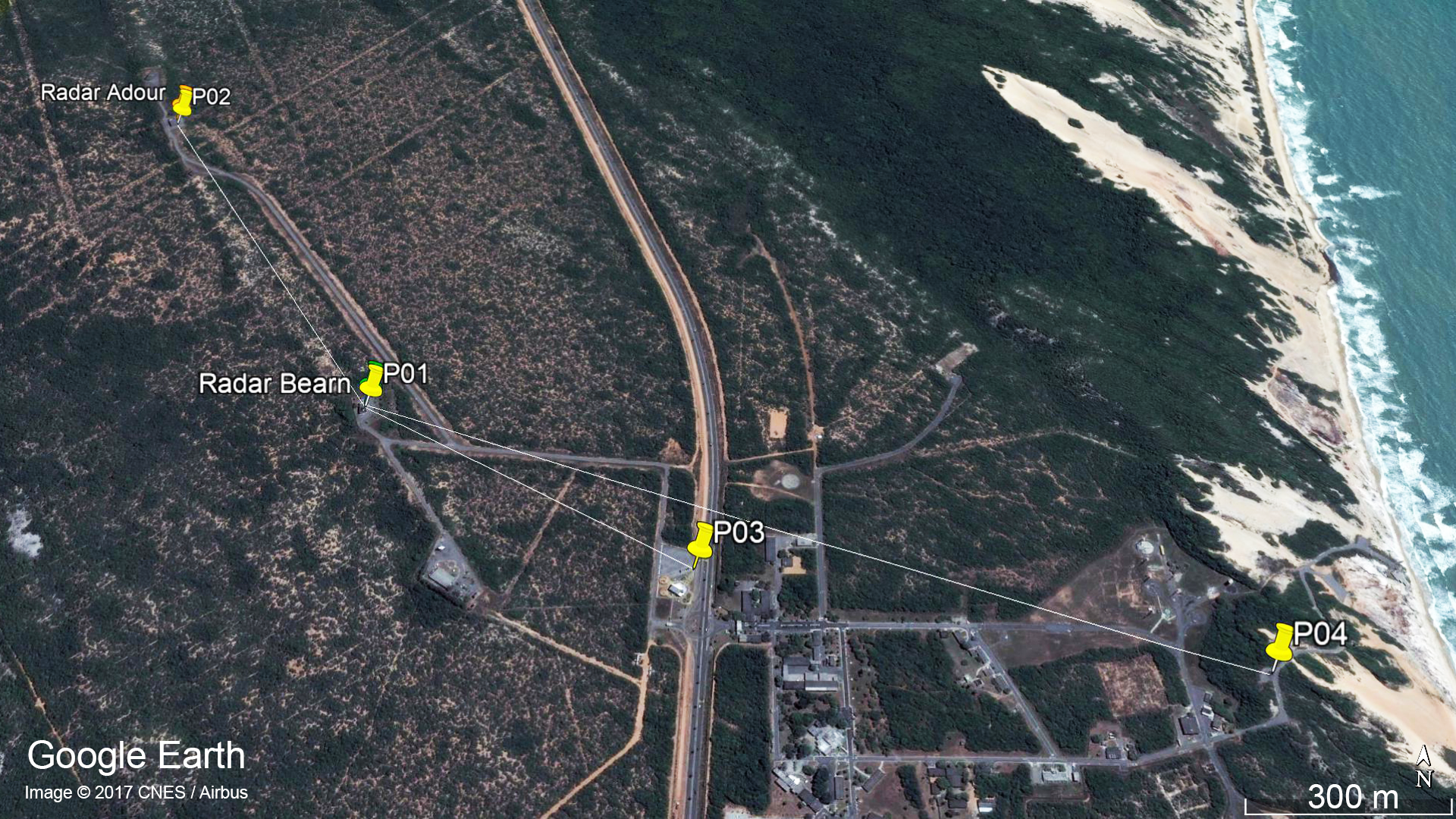}
      \caption{Location 
 of measurement~points.}
     \label{fig_01_CLBI}
\end{figure}

In order to analyze the NIR levels from the Béarn radar, four measurement points are chosen and named P01, P02, P03 and P04. Their positions are shown in Figure~\ref{fig_01_CLBI}. They are 9, 960, 690 and 1570 m away from the radar’s antenna, respectively. To~locate the radar’s antenna and all measurement points, we use the WGS-84 and the SIRGAS 2000 coordinate~systems. 

\begin{figure}[H]
     \includegraphics[width=0.7\linewidth]{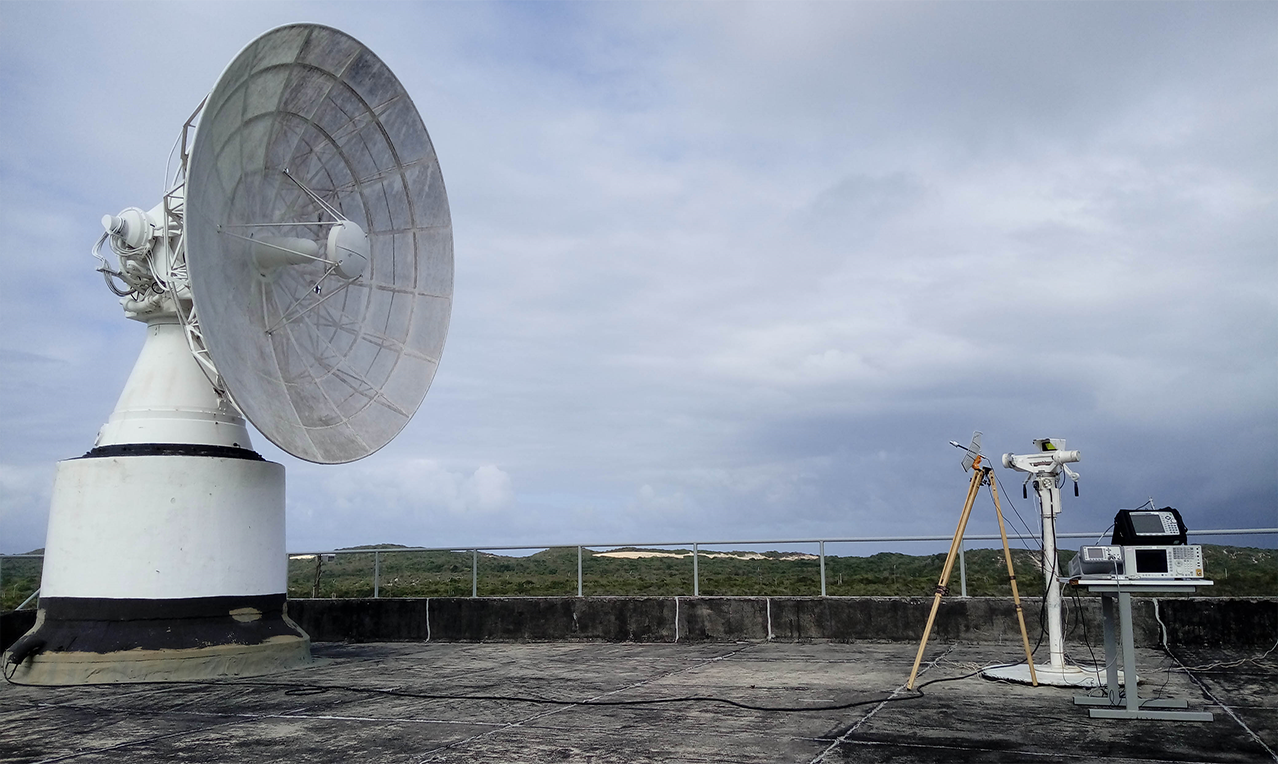}
      \caption{P01 measurement point: Béarn’s antenna on the (\textbf{left}) and its optic designation instrument on the~(\textbf{right}).}
     \label{fig_02_CLBI}
\end{figure}

\begin{figure}[H]
     \includegraphics[width=0.7\linewidth]{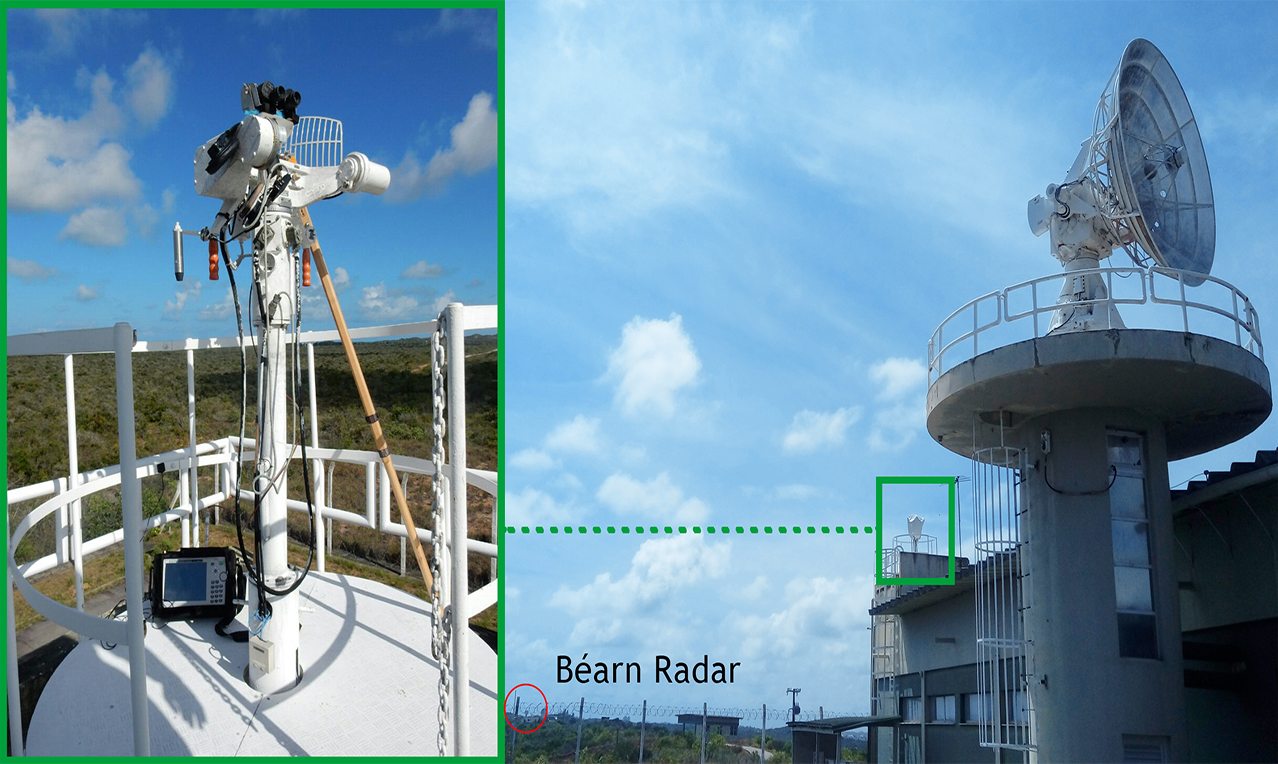}
      \caption{P02 measurement point: Adour’s optic designation instrument on the (\textbf{left}) and its antenna on the~(\textbf{right}).}
     \label{fig_03_CLBI}
\end{figure}

\begin{figure}[H]
     \includegraphics[width=0.7\linewidth]{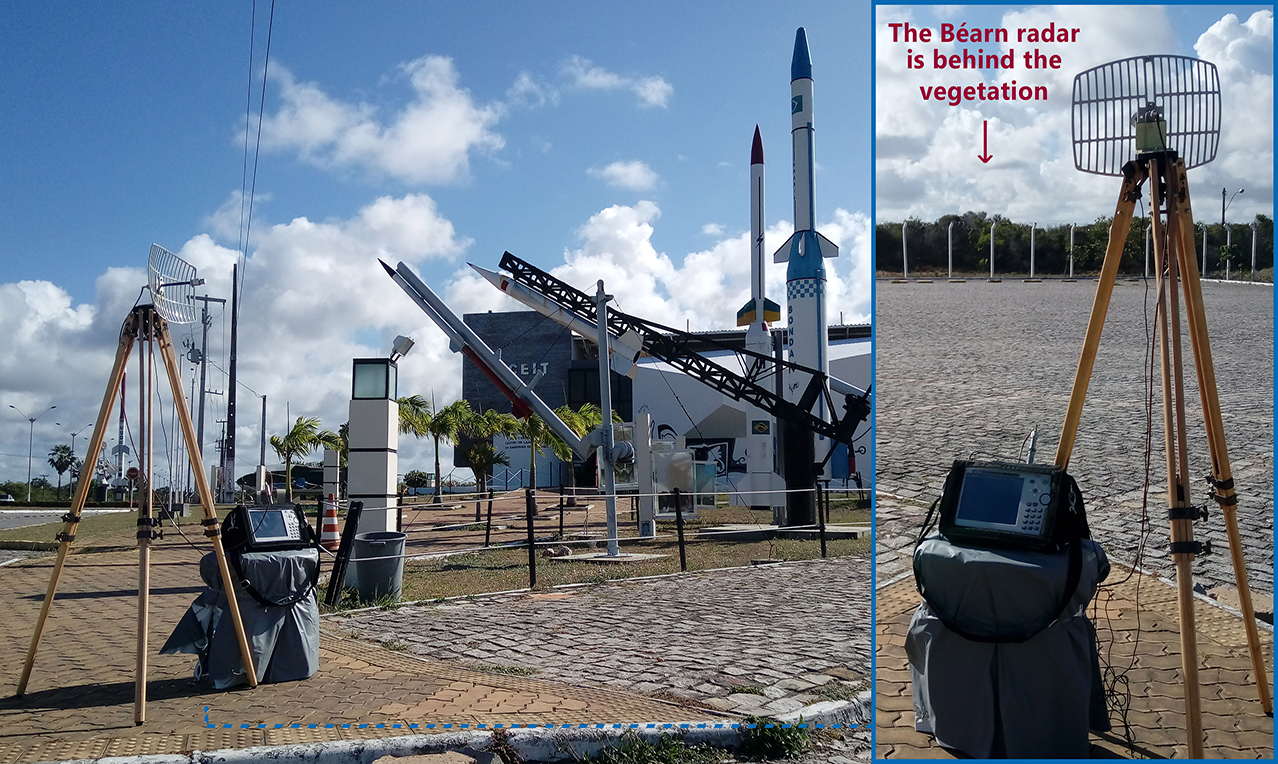}
      \caption{P03 measurement point: entrance of the space~museum.}
     \label{fig_04_CLBI}
\end{figure}

\begin{figure}[H]
     \includegraphics[width=0.75\linewidth]{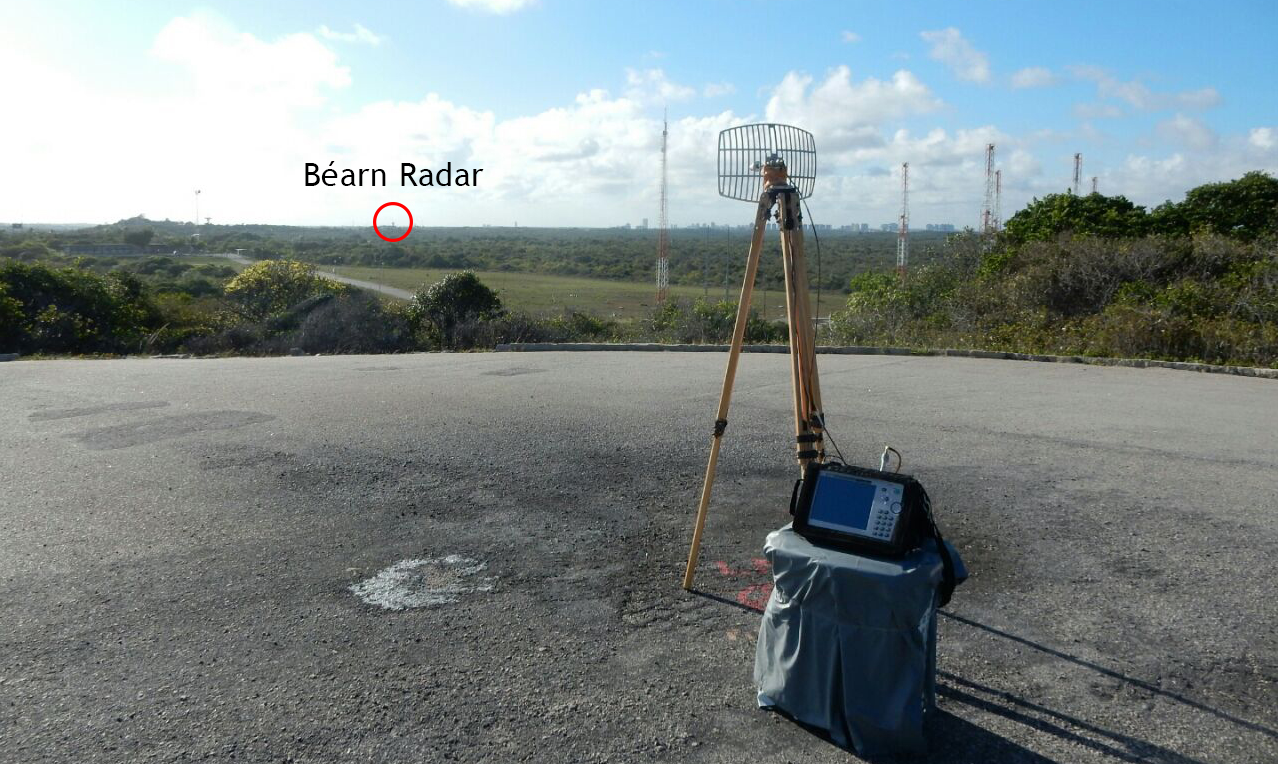}
      \caption{P04 measurement point: mobile launch platform~area.}
     \label{fig_05_CLBI}
\end{figure}

Points P01 and P02 are located at the optic designation instruments of Béarn (Figure~\ref{fig_02_CLBI}) and Adour (Figure~\ref{fig_03_CLBI}) radars, which are operated by professionals during launching campaigns to obtain the aerospace artifact position at its visual flight phase. Point P03 is located at a space museum (Figure~\ref{fig_04_CLBI}), which is one of the most visited places by tourists in the region. The~last point, P04, is located at a mobile launch platform area (Figure~\ref{fig_05_CLBI}), where people work during launching campaigns. These points were selected following two criteria: locations with a high probability of both exposure and presence of~people.

\unskip

\unskip

\unskip

Three of the four measurement points (P01, P02 and P04) have a \rev{LoS} propagation condition with first Fresnel zone clearance, as~shown in Figures~\ref{fig_02_CLBI}, \ref{fig_03_CLBI} and \ref{fig_05_CLBI}. Otherwise, there is no \rev{LoS} for the P03 link due to a shadowing caused by vegetation on the signal path, as~shown in Figure~\ref{fig_04_CLBI}.

All measurement points, with~exception of P01, are considered to be in the far field region of the transmitter, in~which the waves can be considered plane. Point P01, the~closest one to the radar’s antenna, is in the near field region. The~limit distance between the near field and far field can be calculated by \rev{Equation}~(\ref{eq1})~\cite{ANATEL_ATO_458}, where L is the maximum dimension of the radar’s antenna, $\lambda$ is the wavelength, $f$ is the carrier frequency and $c$ is the light speed (3 $\cdot$ 10$^8$ m/s). 

\begin{equation}\label{eq1}
d_{lim}=\frac{2 L^{2}}{\lambda}=\frac{2 f L^{2}}{c}
\end{equation}

Assuming $L$ equals to 4 m~\cite{THOMSON-CSF} and $f$ equals to 5800 MHz, the~calculated near far field limit distance is 618.67 m, which confirms that only P01 is within the near field~region.


\subsection{Measurement~Setup}

\revtech{A spectrum analyzer and a directional antenna (22 dBi of gain) are used to restrict signals from other sources into the frequency and space domains, respectively}. The~spectrum analyzer allows the average power measurement of a specific bandwidth centered on the carrier frequency \rev{of the radar signal, excluding emissions from sources in other frequency bands}. The~directional antenna allows the signal reception to be mainly from the direction of the radar’s antenna. \rev{Furthermore, t}he use of a receiving antenna with such a gain ensures a high dynamic range for the~setup.


An isotropic probe – a quasi-isotropic antenna – is the most used on NIR measurements, because~it can capture signals from all directions and in the three polarization axes, which will eventually compose the total human exposure. In~the specific case of this work, the~possible drawback of not measuring multipath signals due to the high receiving antenna directivity, its fixed linear polarization and cross-polarization discrimination, is avoided since radar half power beam width is extremely narrow \rev{(about 0.92º)} and there are no significant features causing reflections in the scenario. Therefore, it is possible to use a directional antenna to measure the NIR levels exclusively from a trajectography radar, as~proposed. \rev{Anatel establishes the possibility of using either an isotropic or a directional antenna~\cite{ANATEL_ATO_458}.}

\rev{T}o measure the higher NIR levels possible\rev{,} by capturing the main lobe of the radar emission\rev{,} both radar and receiving antennas were precisely directed towards each other for all measurement points, except~at point P01, due to a characteristic of the radar design. Its antenna could not be exactly positioned toward the receiving antenna, as~shown in Figure~\ref{fig_02_CLBI}. There is a difference of 16.43° \rev{in elevation} between the desired position and the lowest possible radar antenna position. Thus, for~the point P01, the~NIR levels were measured from a side lobe of the radar antenna. Since P01 corresponds to the position of the optical designation instrument, this means an operator cannot receive radiation from the antenna’s main~lobe. 


The receiving antenna \rev{was connected to the spectrum analyzer} and was placed at 1.60 m from the ground, assuming to be approximately the mean height of the Brazilian people. Furthermore, the~radar’s antenna and the receiving antenna were set to vertical polarization, and~the measurements were performed for 6 min, resulting in a mean value of power. This measurement period is in accordance with Anatel regulations~\cite{ANATEL_ATO_458}.


\rev{T}he Channel Power function of the spectrum analyzer was used to perform the power measurement of the radar signal. \rev{It} is designed to measure the average power across a given frequency band. \rev{So, it} was set to 5800 MHz of center frequency and to 6 MHz of integrated bandwidth (IBW), encompassing at least the main and the first adjacent sidelobes of the signal spectrum~\cite{Rohde}. \rev{The} average power measurement \rev{was performed} for all points, using the setup shown in Figures~\ref{fig_02_CLBI}--\ref{fig_05_CLBI}. 

\subsection{Bench~Tests}

\begin{figure}[H]
     \includegraphics[width=0.9\linewidth]{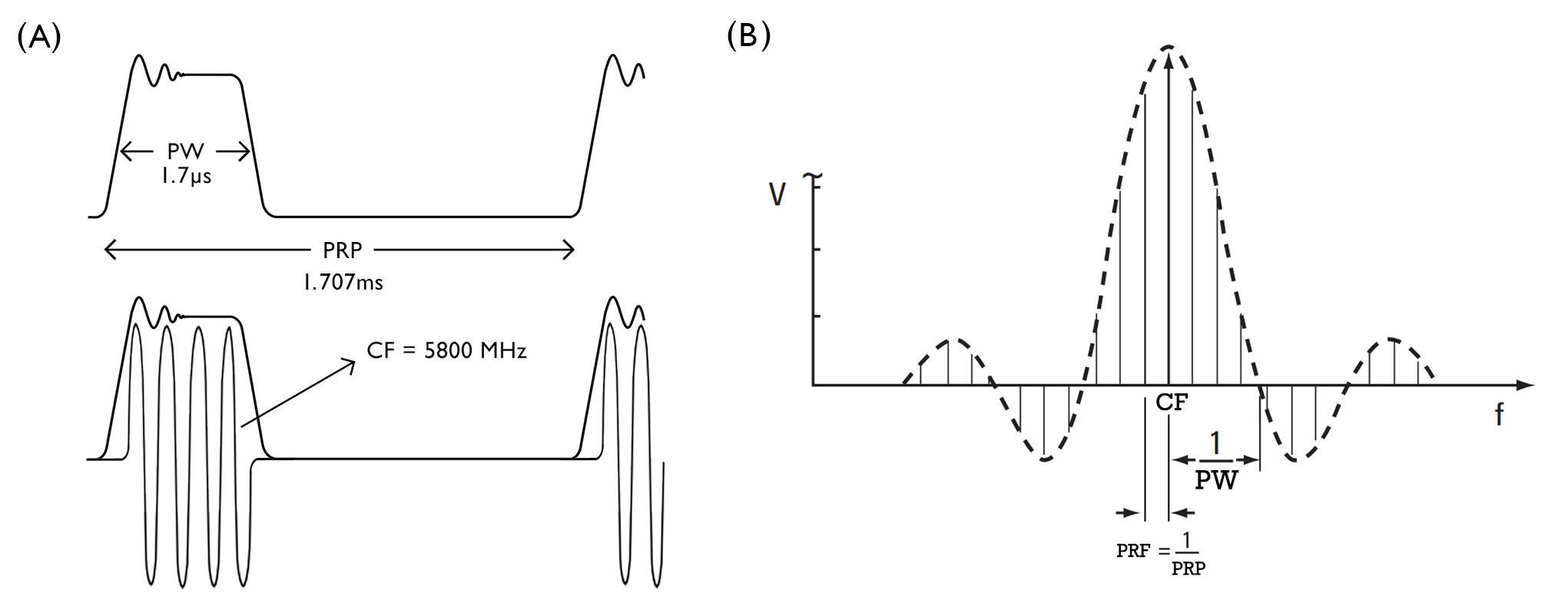}
      \caption{(\textbf{A}) Radar 
 signal in time domain; (\textbf{B}) Radar signal spectrum. Adapted from~\cite{KEYSIGHT}.}
     \label{fig_06_CLBI}
\end{figure}

\rev{Before starting field measurements, some tests} were performed to confirm the\linebreak  specifications of the radar transmitted pulsed signal. In~order to measure the pulse width (PW) and the pulse repetition period (PRP) of the signal, a~Schottky Diode Detector and an oscilloscope connected to the radar transmitter output were used. This measurement confirmed a pulse width of 1.7 µs and a pulse repetition period of 1.707 ms, as~specified in the radar technical specs~\cite{THOMSON-CSF}. Figure~\ref{fig_06_CLBI}A illustrates a representation of the signal pulse in the time domain. A~power meter and both spectrum analyzers connected to the output of the radar transmitter confirmed the transmitted power. Our records measured 0.851 kW of average power (59.3 dBm), a~value 0.7 dB lower than the radar technical specification. This measurement also confirmed the signal carrier frequency of 5800~MHz.

\rev{Tests were also} performed to ensure the reliability of the spectrum analyzers’\linebreak measurements. The~equipment was tested with a simulated signal, produced by an RF signal generator, with~the same characteristics as the radar signal. Its spectrum is illustrated in Figure~\ref{fig_06_CLBI}B. \revtech{Since the signal has a pulsed shape, its frequency spectrum is similar to a sync function. A~power meter was used to confirm the average power measurement of the spectrum analyzers.} 



\unskip


\subsection{Equipment}

All equipment was calibrated before each measurement campaign. Table~\ref{tab_equip} presents the full list of equipment~used.


\begin{table}[H]
    \caption{List 
 of~Equipment.}
    \label{tab_equip}
\setlength{\cellWidtha}{\textwidth/3-2\tabcolsep-1in}
\setlength{\cellWidthb}{\textwidth/3-2\tabcolsep+0.5in}
\setlength{\cellWidthc}{\textwidth/3-2\tabcolsep+0.5in}
		\begin{tabularx}{\textwidth}{>{\raggedright\arraybackslash}m{\cellWidtha}>{\raggedright\arraybackslash}m{\cellWidthb}>{\raggedright\arraybackslash}m{\cellWidthc}}
			\toprule
	  \multicolumn{1}{l}{\footnotesize \textbf{N\textsuperscript{\underline{o}}}} &
	  \multicolumn{1}{l}{\footnotesize \textbf{Item}} &
	  \multicolumn{1}{l}{\footnotesize \textbf{Specification}} \\ \hline
	    \footnotesize 1            & \footnotesize Spectrum Analyzer 1   & \footnotesize Anritsu-MS2720T         \\
	    \footnotesize 2            & \footnotesize Spectrum Analyzer 2   & \footnotesize Agilent-N9010A   \\
	    \footnotesize 3            & \footnotesize RF Generator   & \footnotesize Rohde \& Schwarz-SMR20        \\
	    \footnotesize 4            & \footnotesize Power Meter   & \footnotesize Agilent-N1913A      \\
	    \footnotesize 5            & \footnotesize Power Sensor   & \footnotesize Agilent-N8481A     \\
	    \footnotesize 6            & \footnotesize Oscilloscope   & \footnotesize Agilent-DSO9104A   \\
	    \footnotesize 7            & \footnotesize Schottky Diode Detector   & \footnotesize Agilent-8473C         \\
	    \footnotesize 8            & \footnotesize Directional Antenna   & \footnotesize Hyper Gain-HG5822G, gain: 22~dBi        \\
	    \footnotesize 9            & \footnotesize RF Cables   & {\footnotesize 50 $\Omega$; SMA-SMA; Attenuation: 2.4 dB} \\
	    \footnotesize 10            & \footnotesize Signal Cables   & {\footnotesize 50 $\Omega$; BNC-BNC}\\
	    \footnotesize 11           & \footnotesize Connectors   & \footnotesize Type: SMA-N; Attenuation: 0.05~dB        \\
	    \footnotesize 12            & \footnotesize \rev{Attenuators}   & \footnotesize 10 dB\rev{/20 dB/30 dB}        \\
	    \footnotesize 13            & \footnotesize Wood Tripod  & \footnotesize Maximum Height: 2~m        \\
	\bottomrule
	\end{tabularx}
\end{table}
\unskip

\section{Results and~Discussion}
\label{sec_results}


\revminor{This section presents measurements and prediction results, followed by comparative analyzes regarding the limits established by Anatel and ICNIRP.}

\subsection{\rev{Measurement~Results}}

\begin{figure}[H]
     \includegraphics[width=0.8\linewidth]{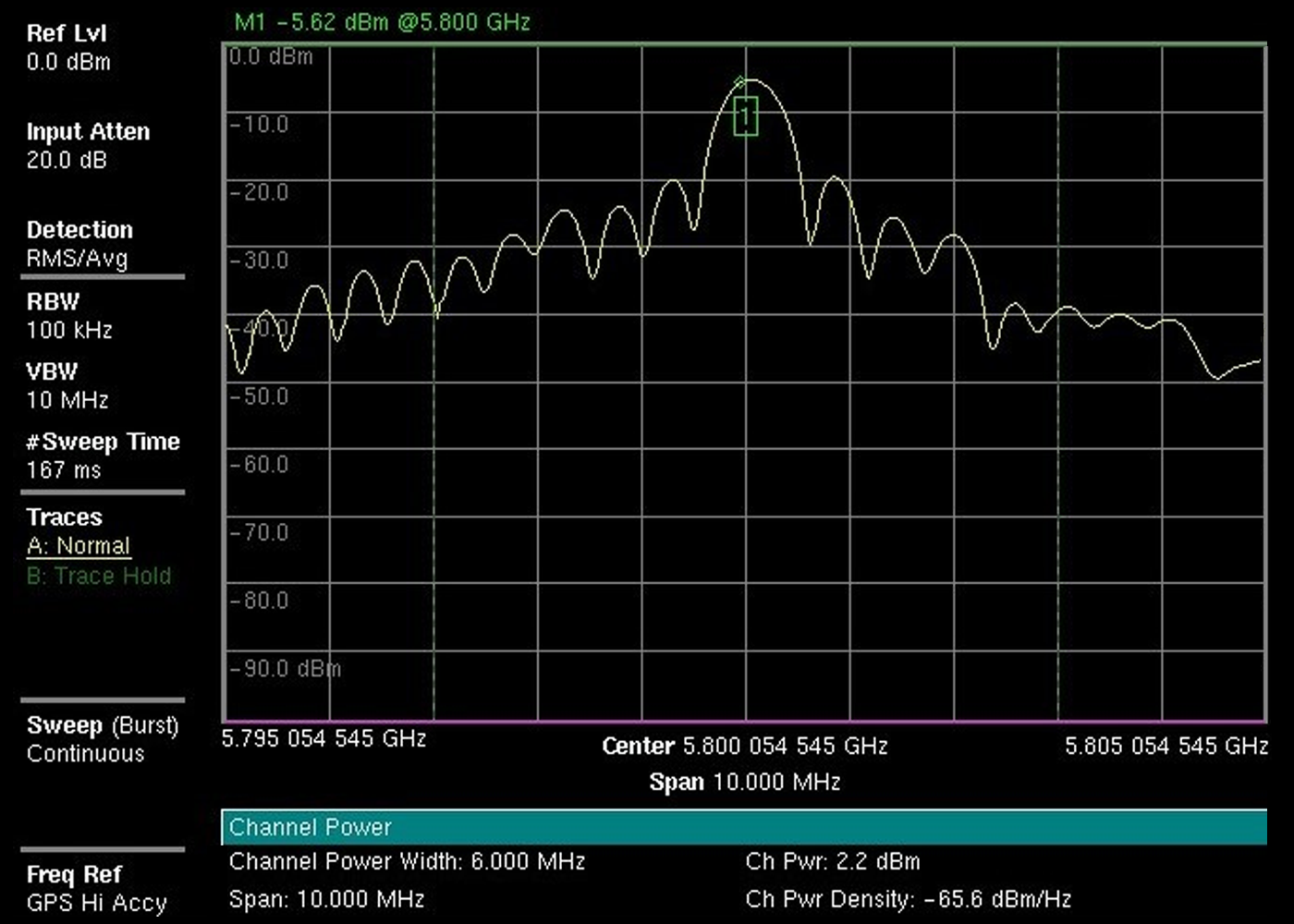}
      \caption{Radar 
 signal spectrum and measured average power (Anritsu MS2720T).}
     \label{fig_07_CLBI}
\end{figure}

By measuring the average power level of the radar signal, it is possible to calculate the power density level of the emission, which has a limiting value as specified by ICNIRP~\cite{ICNIRP-2020} and Anatel~\cite{ANATEL_ATO_458}. Figure~\ref{fig_07_CLBI} shows the spectrum of the received signal at point P01 and its average power value measured by the Anritsu MS2720T spectrum~analyzer. 

From the average power values, we calculate the average power at the terminals of the receiving antenna by using \rev{Equation}~(\ref{eq2}). 

\begin{equation}\label{eq2}
P_{rx}=P_{i}+A_{tt}+P_{off}
\end{equation}

\noindent where $P_{rx}$ is the average power at the terminals of the receiving antenna (dBm), $P_i$ is the average power measured by the instrument (dBm), $A_{tt}$ is the total attenuation between the measurement instrument and the terminals of the receiving antenna (dB) and $P_{off}$ is the instrument power offset (dB). The~calculated values of the average power at the terminals of the receiving antenna are shown in Table~\ref{tab_power}. From~these values, we calculate the power density values by using \rev{Equation}~(\ref{eq3}). 

\begin{equation}\label{eq3}
S=\frac{P_{rx}4\pi}{\lambda^{2}G_{r}}
\end{equation}

\noindent where	$S$ is the power density (W/m$^2$), $P_{rx}$ is the average power at the terminals of the receiving antenna (W), $G_r$ is the receiving antenna gain and $\lambda$ is the signal wavelength (m). The~calculated power density values are also shown in Table~\ref{tab_power}.

\begin{table}[H]
    \caption{Field 
 Power Measurement and Computed Power~density.}
    \label{tab_power}
\newcolumntype{C}{>{\centering\arraybackslash}X}
\begin{tabularx}{\textwidth}{CCCC}
\toprule
	  \multicolumn{1}{l}{\footnotesize \textbf{Point}} &
	  \multicolumn{1}{l}{\footnotesize \textbf{\makecell{Spectrum Analyzer\\ Model}}} &
	  \multicolumn{1}{l}{\footnotesize \textbf{\makecell{Average Power at the \\ Terminals of the Antenna (dBm)}}} &
	  \multicolumn{1}{l}{\footnotesize \textbf{\makecell{Power Density (W/m$^2$)}}} \\ \hline
	    \footnotesize P01            & \footnotesize Agilent - N9010A   & \footnotesize 29.62   &  \footnotesize 27.15    \\
	    \footnotesize P01            & \footnotesize Anritsu - MS2720T   & \footnotesize 29.70   & \footnotesize 27.66   \\
	    \footnotesize P02            & \footnotesize Anritsu - MS2720T   & \footnotesize 16.40   & \footnotesize 1.29    \\
	    \footnotesize P03            & \footnotesize Anritsu - MS2720T   & \footnotesize 4.60   & \footnotesize  0.09   \\
	    \footnotesize P04            & \footnotesize Anritsu - MS2720T   & \footnotesize 11.50   & \footnotesize 0.42   \\
\bottomrule
\end{tabularx}
\end{table}
\unskip

\subsection{\rev{Predictions}}

\rev{The power density values} for the points in the far field region (P02, P03 and P04) \rev{were predicted} using \rev{Equation}~(\ref{eq4})~\cite{C95-2002}, which assumes free space propagation.  
\begin{equation}\label{eq4}
S=\frac{P_{t}G_{t}}{4 \pi d^{2}}
\end{equation}

\noindent where $P_t$ is the transmitted average power, $G_t$ is the transmitting antenna gain and $d$ is the distance from the measurement point to the transmitting antenna. We assume a transmitter average power of 0.851 kW and a transmitter antenna gain of 44~dBi.


Equation~\rev{(}\ref{eq5}\rev{)}~\cite{C95-2002, mum1961}, which evaluates the maximum power density expected for the near field, was used for point P01, located in this~region.


\begin{equation}\label{eq5}
S_{m}=\frac{4P_{t}}{A}
\end{equation}

\noindent where $S_m$  is the maximum power density (W/m$^2$) and $A$ is the area of the transmitting antenna (m$^2$). Equation~\rev{(}\ref{eq5}\rev{)} provides the maximum power density that can exist on the axis of the antenna beam that is focused at infinity, with~no reflections. However, if~the calculated prediction reveals a power density value equal or greater than the maximum permissible exposure (MPE), we must assume that this value may exist at any point in the near field region~\cite{C95-2002}. The~predicted and measured values for all four points are shown in Table~\ref{tab_power_predit}. 


\begin{table}[H]
    \caption{Predicted 
 and Measured Power~Density.}
    \label{tab_power_predit}
\newcolumntype{C}{>{\centering\arraybackslash}X}
\begin{tabularx}{\textwidth}{CCCC}
\toprule
	  \multicolumn{1}{l}{\footnotesize \textbf{Point}} &
	  \multicolumn{1}{l}{\footnotesize \textbf{\makecell{Predicted Power Density \\(W/m$^2$)}}} &
	  \multicolumn{1}{l}{\footnotesize \textbf{\makecell{Measured Power Density  \\ (W/m$^2$)}}} &
	  \multicolumn{1}{l}{\footnotesize \textbf{\makecell{Anatel / ICNIRP Limits\\ (W/m$^2$)}}} \\ \hline
	    \footnotesize P01            & \footnotesize \rev{270.92}   & \footnotesize 27.66   &  \footnotesize 10 */50 **    \\
	    \footnotesize P02            & \footnotesize 1.31   & \footnotesize 1.29   & \footnotesize 10 */50 **    \\
	    \footnotesize P03            & \footnotesize 2.53   & \footnotesize 0.09   & \footnotesize  10 */50 **   \\
	    \footnotesize P04            & \footnotesize 0.49   & \footnotesize 0.42   & \footnotesize 10 */50 **  \\
	\bottomrule
	\end{tabularx}
\noindent\footnotesize{* General limit/** Occupational Limit.}
\end{table}
\unskip

\subsection{\rev{Analysis and~Discussion}}

\revminor{Comparative analyzes between measurements and predictions and between measurements and limits established by Anatel and ICNIRP are presented here for each measurement point.}

\rev{\subsubsection{Measurements and~Predictions}}
Comparing the predicted and measured power density values for point P01, the~power density measured value is considerably lower. We advocate the coherence of these results because the measurement was taken in a point (P01) misaligned to the antenna axis and the predicted value represents the maximum possible power density level that occurs at the axis of the antenna~beam. 

In contrast, the~predicted and measured values are very close for points P02 and P04. This result is also reasonable since they are in a \rev{LoS} condition scenario, with~no additional meaningful attenuation other than free space, a~consequence of a static link configured by a directional receiving antenna and a very narrow beam transmitting~antenna. 

Otherwise, for~the P03 point, the~measured value is 96.44\% far from the predicted value. This result is explained by the shadowing caused by vegetation located on the signal path, configuring a \rev{N}on-\rev{L}ine of \rev{S}ight \rev{(NLoS)} scenario condition, as~shown in Figure~\ref{fig_04_CLBI}.

\subsubsection{\rev{Measurements and~Limits}}

Comparing the results of measured power density with the ICNIRP’s~\cite{ICNIRP-2020} and \linebreak Anatel’s~\cite{ANATEL_ATO_458} limits (Table~\ref{tab_pop_exp} or the last column of Table~\ref{tab_power_predit}), the~values for all measurement points are below the limit for occupational public (50 W/m$^2$). However, for~general population power density limit (10 W/m$^2$), the~measured value at P01 is above the limit. It evidences that the general population cannot have access to the P01 area. In~addition,  as~stated by regulations, the~occupational public needs to be warned about the non-ionizing radiation risk at point P01 and the permanence time at this area needs to be controlled when the radar is in~operation.



\section{Conclusions}
\label{sec_conclu}

\revtech{This paper presents an NIR measurement campaign using a specific method for \linebreak trajectography radars. We perform and analyze the power density measurements from a C-band trajectography radar considering the radar's high gain and locations with a high probability of both exposure and presence of people.}

\revtech{The main contributions of this work are summarized below.}
\begin{enumerate}
\item \revtech{A specific method for measuring non-ionizing radiation power density levels due to trajectography radars is proposed. To~the best of our knowledge, there are no studies about it};
\item \revtech{Unpublished NIR measurement results are obtained;}
\item \revtech{The need to adopt control and alert measures was verified due to the high NIR level (above the limit for the general population) at a point close to the radar antenna.}
\end{enumerate}

\rev{The} measurement setup is able to characterize the NIR exclusively from the radar emissions, by~using the spectrum analyzer channel power function and a directional receiving antenna. Four measurement points were selected, three at the far field region and one at the near-field region of the radar’s~antenna. 

All measured and predicted values were presented and discussed, considering each measurement scenario. As~the main qualitative results, we concluded that all measured power density values are below the occupational population limit established by ICNIRP and Anatel. However, the~general population limit is exceeded at one point located near the radar antenna. It shows the need for restricting general population access to the area very close to the Béarn’s antenna as well as the need for preventive actions to avoid health injuries to the occupational population, by~giving orientation about the risks and controlling the permanence time at the area around the radar’s antenna (actions already performed by CLBI regarding its radars).


\vspace{6pt} 

\authorcontributions{Conceptualization and methodology: J.M.L.B.F. and M.E.C.R.; software and data curation: J.M.L.B.F., M.M.d.M.C., D.L.F. and W.S.A.; validation: M.E.C.R. and V.A.d.S.J.; formal analysis and investigation: J.M.L.B.F., M.E.C.R. and V.A.d.S.J.; resources: J.M.L.B.F. and A.G.D.; writing---original draft preparation: J.M.L.B.F. and M.M.d.M.C.; writing---review, editing and visualization: J.M.L.B.F., M.E.C.R. and V.A.d.S.J.; supervision: M.E.C.R. and V.A.d.S.J. All authors have read and agreed to the published version of the~manuscript.}


\funding{This study was financed in part by the Coordena\c{c}\~{a}o de Aperfei\c{c}oamento de Pessoal de N\'{i}vel Superior - Brasil (CAPES) - Finance Code 001.}

\institutionalreview{Not applicable. }

\informedconsent{Not applicable.} 

\dataavailability{Not applicable. } 

\acknowledgments{The authors would like to thank the Centro de Lançamento da Barreira do Inferno of Brazilian Air Force for the partnership. Special thanks to the Béarn radar’s team and to the members of the research and innovation sector of~CLBI. } 

\conflictsofinterest{The authors declare no conflict of~interest.} 


\begin{adjustwidth}{-\extralength}{0cm}

\reftitle{References}

\end{adjustwidth}

\begin{thebibliography}{999}

\bibitem[Lacomme \em{et~al.}(2001)Lacomme, Hardange, Marchais, and
  Normant]{lac2001}
Lacomme, P.; Hardange, J.P.; Marchais, J.C.; Normant, E.
\newblock {\em {Air and Spaceborne Radar Systems: An Introduction}}, 1st ed.;
  William Andrew Publishing: Norwich, NY, USA,  2001; pp. 249--251.
\newblock {ISBN} 9781891121135.

\bibitem[THOMSON-CSF(1964)]{THOMSON-CSF}
THOMSON-CSF.
\newblock \emph{Béarn {Radar Technical Order, Paris}}; THOMSON-CSF: La Défense, France,  1964. 


\bibitem[ICNIRP(2020)]{ICNIRP-2020}
ICNIRP.
\newblock {International Commission on Non-Ionizing Radiation Protection,
  ICNIRP Guidelines for Limiting Exposure to Electromagnetic Fields (100 kHz to
  300 GHz)}.
\newblock {\em Health Phys.} {\bf 2020}, {\em 118},~483–524.
\newblock https://doi.org/10.1097/HP.0000000000001210.

\bibitem[Anatel(2019)]{ANATEL_ATO_458}
Anatel.
\newblock \emph{Agência Nacional de Telecomunicações (Anatel), Ato nº458}; Anatel: Brasília, Brazil, 
  2019.



\bibitem[Bhargava and Rattanadecho(2022)]{bha2022}
Bhargava, D.; Rattanadecho, P.
\newblock Microstrip Antenna for Radar-Based Microwave Imaging of Breast
  Cancer: Simulation Analysis.
\newblock {\em Int. J. Commun. Antenna Propag.}
  {\bf 2022}, {\em 12},~47--53.

\bibitem[Meng \em{et~al.}(2020)Meng, Lin, Zang, Qing, and Nikolova]{men2020}
Meng, Y.; Lin, C.; Zang, J.; Qing, A.; Nikolova, N.K.
\newblock Ka Band Holographic Imaging System Based on Linear Frequency
  Modulation Radar.
\newblock {\em Sensors} {\bf 2020}, {\em 20},~6527.

\bibitem[Owda \em{et~al.}(2020)Owda, Owda, and Rezgui]{owd2020}
Owda, A.Y.; Owda, M.; Rezgui, N.D.
\newblock Synthetic aperture radar imaging for burn wounds diagnostics.
\newblock {\em Sensors} {\bf 2020}, {\em 20},~847.

\bibitem[Peleg \em{et~al.}(2018)Peleg, Nativ, and Richter]{pel2018}
Peleg, M.; Nativ, O.; Richter, E.D.
\newblock Radio frequency radiation-related cancer: Assessing causation in the
  occupational/military setting.
\newblock {\em Environ. Res.} {\bf 2018}, {\em 163},~123--133.

\bibitem[Yakymenko \em{et~al.}(2011)Yakymenko, Sidorik, Kyrylenko, and
  Chekhun]{yak2011}
Yakymenko, I.; Sidorik, E.; Kyrylenko, S.; Chekhun, V.
\newblock{ Long-term exposure to microwave radiation provokes cancer growth:
  Evidences from radars and mobile communication systems. \emph{Exp. Oncol.} {\textbf{2011}}, } 33(2), 62--70
.

\bibitem[Goldsmith(1995)]{gol1995}
Goldsmith, J.R.
\newblock Epidemiologic evidence of radiofrequency radiation (microwave)
  effects on health in military, broadcasting, and occupational studies.
\newblock {\em Int. J. Occup. Environ. Health}
  {\bf 1995}, {\em 1},~47--57.

\bibitem[Joseph \em{et~al.}(2012)Joseph, Goeminne, Vermeeren, Verloock, and
  Martens]{jos2012}
Joseph, W.; Goeminne, F.; Vermeeren, G.; Verloock, L.; Martens, L.
\newblock Occupational and public field exposure from communication,
  navigation, and radar systems used for air traffic control.
\newblock {\em Health Phys.} {\bf 2012}, {\em 103},~750--762.

\bibitem[Martin \em{et~al.}(2013)Martin, Alves, and Gomes]{mar2013}
Martin, I.M.; Alves, M.A.; Gomes, M.P.
\newblock The electromagnetic spectrum from 1 Hz to 9.4 GHz near ground level
  in the region of S{\~a}o Jos{\'e} dos Campos, SP, Brazil.
\newblock In Proceedings of the  2013 SBMO/IEEE MTT-S International Microwave \& Optoelectronics
  Conference (IMOC),  Rio de Janeiro, Brazil, 4--7 August 2013; 
 pp. 1--4.

\bibitem[Halgamuge(2015)]{hal2015}
Halgamuge, M.N.
\newblock {Radio hazard safety assessment for marine ship transmitters:
  measurements using a new data collection method and comparison with ICNIRP
  and ARPANSA Limits}.
\newblock {\em Int. J. Environ. Res. Public Health} {\bf 2015}, {\em 12},~5338--5354.
\newblock https://doi.org/10.3390/ijerph120505338.

\bibitem[Wollinger(2003)]{wol2003}
Wollinger, P.R.
\newblock Estudo dos Níveis de Radiação Eletromagnética em Ambiente Urbano.
\newblock Master's Thesis, Universidade Federal de Santa Catarina (UFSC),
  Florianópolis, SC, Brazil, 2003.

\bibitem[WHO()]{WHO_EMF}
WHO.
\newblock {World Health Organization (WHO)---International EMF Project}. Available online: \url{https://www.who.int/initiatives/the-international-emf-project} (accessed on  13 October 2021).


\bibitem[ICNIRP(1998)]{ICNIRP-1998}
ICNIRP.
\newblock {International Commission on Non-Ionizing Radiation Protection,
  ICNIRP Guidelines for Limiting Exposure to Time-Varying Electric, Magnetic
  and Electromagnetic Fields (up to 300 GHz)}.
\newblock {\em Health Phys.} {\bf 1998}, {\em 74},~494--522.
\newblock {ISBN} 978-3- 9804789-6-0.

\bibitem[de~F.~Diniz \em{et~al.}(2021)de~F.~Diniz, de~Sousa~Jr., Rodrigues,
  Mendonça, da~Silva, and Pinheiro]{jmoe-2021}
de~F.~Diniz, A.B.; de~Sousa~Jr., V.A.; Rodrigues, M.E.C.; Mendonça, H.B.;
  da~Silva, G.S.; Pinheiro, F.S.R.
\newblock {Non-Ionizing Radiation Analysis in Close Proximity to Antenna Tower:
  A Case Study in Northeast Brazil}.
\newblock {\em J. Microwaves Optoelectron. Electromagn. Appl.} {\bf 2021}, {\em 20},~126–142.
\newblock https://doi.org/10.1590/2179-10742021v20i1833.

\bibitem[Pinheiro \em{et~al.}(2015)Pinheiro, de~Oliveira~Maranhão, Filho,
  da~Costa~Rodrigues, da~Silva, de~Sousa, Sanchis, Câmara, da~Silva~Gonçalo,
  and de~Abiahy Carneiro~da Cunha~Braga]{Marcio_2015}
Pinheiro, F.S.R.; de~Oliveira~Maranhão, T.M.; Filho, M.B.; da~Costa~Rodrigues,
  M.E.; da~Silva, G.S.; de~Sousa, T.P.; Sanchis, M.A.B.; Câmara, A.L.S.;
  da~Silva~Gonçalo, J.P.; de~Abiahy Carneiro~da Cunha~Braga, A.
\newblock {Assessment of non-ionizing radiation from radio frequency energy
  emitters in the urban area of Natal City}.
\newblock  \emph{Sci. Res. Essays}  \textbf{2015}, \emph{2}, 79--85.
\newblock  https://doi.org/10.1109/IMOC.2013.6646602.

\bibitem[Rodrigues \em{et~al.}(2013)Rodrigues, Pinheiro, Braga, Sousa,
  Gonçalo, Sanchis, and Câmara]{Marcio_2013}
Rodrigues, M.E.C.; Pinheiro, F.S.R.; Braga, A.A.C.C.; Sousa, T.P.; Gonçalo,
  J.P.S.; Sanchis, M.A.B.; Câmara, A.L.S.
\newblock Measurements of non-ionizing radiation on urban environment and
  preliminary assessment of relative contribution among different services.
\newblock In Proceedings of the  2013 SBMO/IEEE MTT-S International Microwave \& Optoelectronics
  Conference (IMOC), Rio de Janeiro, Brazil,  4--7 August 2013; pp. 1--4.

\bibitem[Schwarz(2003)]{Rohde}
Schwarz, R.
\newblock {\emph{Power Measurement on Pulsed Signals with Spectrum Analyzers};
  Application Note;  Rohde \& Schwarz Regional Headquarters Singapore Pte Ltd.: Singapore, 2003.} 


\bibitem[Keysight(2014)]{KEYSIGHT}
Keysight.
\newblock {\emph{Radar Measurements}; Application Note; Keysight: Santa Rosa, CA, USA,  2014.} 



\bibitem[IEEE(2002)]{C95-2002}
\emph{IEEE Standard C95.3-2002};
Recommended Practice for Measurements and
  Computations of Radio Frequency Electromagnetic Fields with Respect to Human
  Exposure to Such Fields, 100 kHz-300 GHz.  IEEE: Piscataway, NJ, USA, 2002.

\bibitem[Mumford(1961)]{mum1961}
Mumford, W.
\newblock Some technical aspects of microwave radiation hazards.
\newblock {\em Proc. IRE} {\bf 1961}, {\em 49},~427--447.

\end{thebibliography}
\end{document}